# Usability of the Size, Spacing, and Depth of Virtual Buttons on Head-Mounted Displays

Kyudong Park[a], Dohyeon Kim[b], Sung H. Han[b*]

[a]Department of Creative IT Engineering, Pohang University of Science and Technology (POSTECH), Pohang, South Korea

[b]Department of Industrial and Management Engineering, Pohang University of Science and Technology (POSTECH), Pohang, South Korea

## Abstract

Virtual reality (VR) allows users to see and manipulate virtual scenes and items through input devices, like head-mounted displays. In this study, the effects of button size, spacing, and depth on the usability of virtual buttons in VR environments were investigated. Task completion time, number of errors, and subjective preferences were collected to test different levels of the button size, spacing, and depth. The experiment was conducted in a desktop setting with Oculus Rift and Leap motion. A total of 18 subjects performed a button selection task. The optimal levels of button size and spacing within the experimental conditions are 25 mm and between 5 mm and 9 mm, respectively. Button sizes of 15 mm with 1-mm spacing were too small to be used in VR environments. A trend of decreasing task completion time and the number of errors was observed as button size and spacing increased. However, large size and spacing may cause fatigue, due to continuous extension of the arms. For depth effects, the touch method took a shorter task completion time. However, the push method recorded a smaller number of errors, owing to the visual push-feedback. In this paper, we discuss advantages and disadvantages in detail. The results can be applied to many different application areas with VR HMD.

# 1. Introduction

Virtual reality (VR) allows users to see and manipulate virtual scenes and items via input devices (Mazuryk and Gervautz, 1996; Burdea and Coiffet, 2003). An American scientist, Ivan Sunderland, first pushed the idea of virtual display technology in 1965. He claimed, "A display connected to a digital computer gives us a chance to gain familiarity with concepts not realizable in the physical world" (Sutherland, 1965). VR technologies enable users to have interactive experiences with its engaging and intuitive nature. It even can captivate novices who are not familiar with new technologies (Zhang, 2017). As VR has consistently developed, the field of introducing VR technologies is progressively expanding, such as computer game (Martel and Muldner, 2017), virtual training (Bertram et al., 2015), automotive industry (Lawson et al., 2016), and healthcare (Rose et al., 2018).

To have an immersive and engaging VR experience, input and output devices are required such as keyboard, data glove, and head-mounted display (HMD)). Among the options, the HMD is the most commonly used device in VR environments. The HMD, like Oculus Rift and Google Cardboard, provide intuitive and immersive experiences within a virtual environment. However, the VR HMD presents difficulties for traditional controlling devices, like physical keyboards (Kim and Kim, 2004). Therefore, indirect methods of manipulation, whereby VR interaction is presented by wielding a tool, are necessary alternatives to overcome the difficulties. These tools provide users more immersive interactions with VR (Sekuler and Blake, 1985). The Samsung HMD Odyssey, for example, is an external controller manufactured for VR environments. On the other hand, direct method of manipulation uses the human body to interact with the virtual world. This method has advantages over the indirect method (Mine et al., 1997). According to their results, the direct methods of manipulation enabled higher performance than the manipulation of virtual objects with users' hands. It provided a more immersive virtual environment to users (Lubos et al., 2014). The direct method using human body is known to give a high focus and impression (IJsselsteijn et al., 2008), which are aspects that have a great influence on usability (Sun et al., 1995). Many studies have followed, studying the direct manipulation of virtual objects, owing to the potential applications in many different fields (Hofmann et al., 2013).

This study aims to understand hand input, a direct method, with VR HMDs and to enhance their use of buttons. To achieve these purposes, an experiment was conducted to investigate the effects of button size, spacing, and depth on the usability with hand interaction. The paper is structured as follows. Section 2 demonstrates the method and Section 3 reports the results of the experiment. Then, key implications and limitations are discussed in Section 4.

## 2. Method

### 2.1. Subjects

A total of 18 subjects (9 males and 9 females) voluntarily participated in the experiment. Their ages ranged from 18 to 27 years (mean = 23.56, SD = 2.59). The subjects were all right-handed. They had normal or corrected vision and freely moved their right hands. Eight had no experience with VR devices, whereas others had an average week's experience.

### 2.2. Experimental design

The experiment used a three-factor (i.e., button size, button spacing, and button depth) within-subjects design. A balanced Latin Square was used to avoid transfer, learning, and fatigue effects.

The independent variables in this experiment were button size (i.e., SIZE, 3 levels), button spacing (i.e., SPACING, 3 levels), and button depth (i.e., METHOD, 2 levels). Owing to the lack of literature related to button selection tasks in VR environments, the level of independent variables was determined based on studies related to the use of buttons on Kiosk-type front-screen devices (Scott and Conzola, 1997; Colle and Hiszem, 2004). Research related to the use of buttons on smart phones and personal digital assistants was also considered (Park and Han, 2010; Parhi et al., 2006; Park, 2010).

The button sizes used in Kiosk-type front-screens were 16 mm, 18 mm, and 20 mm, and there were no time differences in input speed. Colle and Hiszem (2004) found that, among standard button sizes (i.e., 10 mm, 15 mm, 20 mm, and 25 mm), the 20-mm buttons were preferred. Several studies on the use of smart phone buttons suggested that their sizes should be between 7 mm and 10 mm (Park and Han, 2010; Parhi et al., 2006; Park, 2010). In our experiment, a virtual board containing virtual buttons was located 500 mm away from the users' eyes. The distance from a smart phone to user's eyes is normally 322 mm to 362 mm (Bababekova et al., 2011). Thus, the button sizes in our experimental setting were obtained proportionally, resulting in sizes ranging from 10 mm to 15 mm. We conducted a pilot experiment prior to the main experiment, finding that there exists low accuracy of hand movement in VR. Therefore, we increased the button sizes slightly and set the three size levels to 15 mm, 20 mm, and 25 mm. These sizes were appropriate enough for subjects to sense the differences in button size.

The level of spacing was also set based on the button-size studies on Kiosk-type front-screens and smart phones. Scott and Conzola (1997) claimed that spacing levels of 0 mm, 2 mm, and 4 mm were not statistically different. They argued that, as the level of spacing increased, there existed a trend of increasing errors. Alternatively, Colle and Hiszem (2004) found that the spacing of 1 mm and 3 mm was not statistically different. For smart phones, when the button sizes were less than 4 mm, the spacing larger than 2 mm was recommended (Park, 2010). Therefore, the level of spacing in our experiment was set to 1 mm, 5 mm, and 9 mm. In our pilot experiment, this spacing condition was appropriate, enabling users to sense the differences among the three levels.

For button depth, there were two levels of touch and push methods. The depth of the touch method was 0 mm. The touch button sensed the event just as the virtual hand touched the button. In contrast, the push method required users to push and move virtual buttons to a certain threshold. The push-type virtual button sensed its event trigger when the given depth was reached. In the pilot test, the subjects were unable to recognize that the button had a depth requirement of 10 mm. Therefore, the depth of button was reset to 15 mm, so that it would be visually and clearly recognizable.

*2.3. Dependent measures*

Two types of dependent measures (i.e., pressing performance and subjective preference) were collected from the experiment. The pressing performance measures were time-related measures (i.e., task completion time, TIME) and accuracy measures (i.e., number of errors, ERROR). TIME was defined as the time required to complete the task of correctly pressing a target button. Trials with at least one error were eliminated from the data set as in a touch-key design study (Kim et al., 2014), because the error event caused increases of time. ERROR was defined as the number of incorrect button-pressing. Subjective preference (i.e., PREF) was evaluated by subjects on each experimental condition, scoring between 0 and 100 points.

*2.4. Apparatus*

The experiment was conducted using Oculus Rift, a VR HMD developed by Oculus VR. This device has an OLED display, 1,080 × 1,200 resolution per eye, a 90-Hz refresh rate, and 110° field of view. To detect hand motions, Leap motion was attached to the front of the VR device. The prototype used for conducting the selection task was implemented using Unity and C# programming languages. Virtual buttons were placed on a virtual board. The distance from the board and the users' eyes was 500 mm. This value was derived based on a Korean body size survey conducted in 2015 by Statistics Korea. The arm length of the twenties is 583 mm on average. This value was adjusted in our pilot test to 500 mm so that subjects could reach out their arms with ease. Additionally, the size of the virtual hand was set to 177 mm. This value was also based on the Korean body size survey in 2015. The size of a hand is measured from the lowest part of hand to the tip of the middle finger.

The experimental environment is shown in Fig. 1. An instructor and a subject were seated together on the same side of a table. Two sensors were placed on the left and right sides to sense the subjects' HMD, onto which, Leap motion was connected via a desktop computer located in front of the instructor. The instructor constantly monitored the entire experimental process through the monitor.

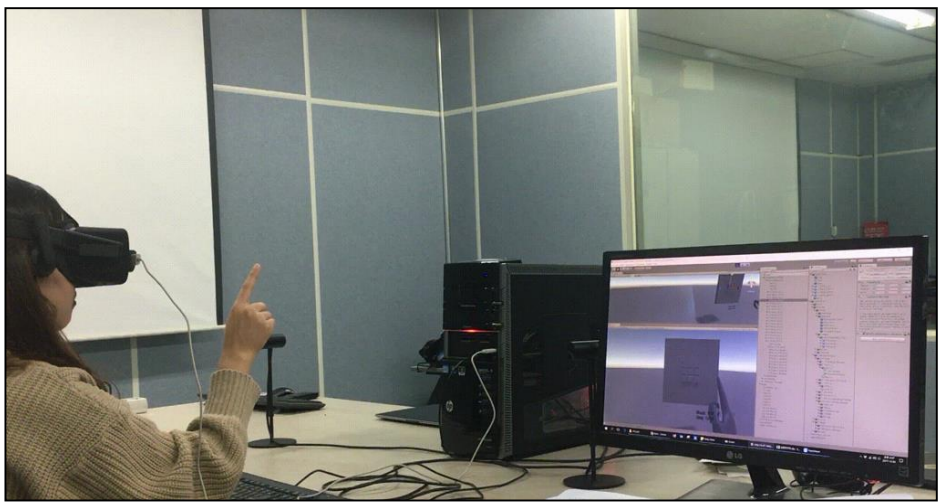

Fig. 1. Experimental environment

## 2.5. Experimental procedure

Subjects completed a demographic questionnaire prior to the experimental task. The demographic data included the individual's age, VR experience, and gestural device experience. Each subject was given written instructions on experimental objectives and procedures. The subjects wore the HMD on their heads and performed practice trials to become familiar with the experimental environment and the virtual buttons. The experiment consisted of 18 conditions. After finishing each condition, subjects were asked to rate their subjective preference on button settings. When the entire experiment was finished, an interview session was conducted. We asked subjects to provide their personal opinions on the size, spacing, and depth of the buttons, as well as any inconvenient experiences during the experiment.

## 2.6. Experimental task

Subjects were presented with nine grey buttons in a 3 × 3 layout (see Fig. 2). When subjects pressed the space button on the keyboard with their left hand, the color of one randomly determined button changed to blue. In the main experiment, the subjects were instructed to select the blue button as accurately as possible. In this experiment, only visual feedback was provided without any acoustic or tactile feedback. The color of the button changes back to its original grey color when touched or pushed to the threshold limit by the subject's index finger. Touching or pushing buttons other than the target button was recorded as an error. The task continued until the subjects correctly touched or pushed the target button. If the blue button was not lit on the board, the subjects could rest. When the subject was ready for the next task, he or she would press the space bar to continue. For one experimental condition, buttons on the board were required to be touched or pushed twice. A total of 18 selection tasks were needed for one experimental condition. Therefore, a single subject performed 324 selection tasks in a single experiment.

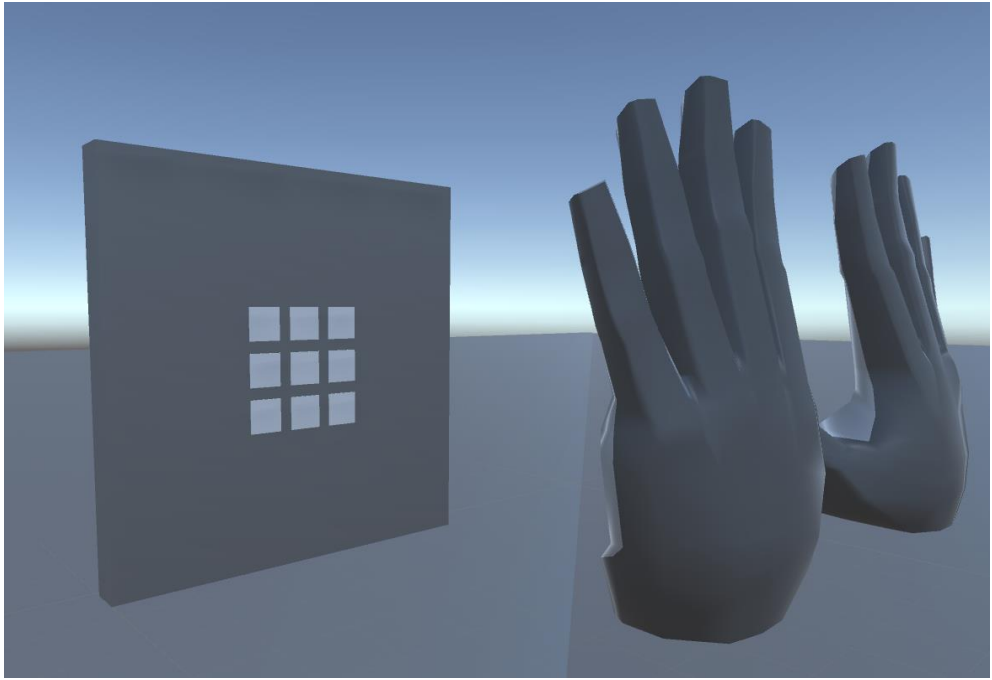

Fig. 2. 3 x 3 button layout for the experiments

## 3. Results

An analysis of variances (ANOVA) was performed. For significant main effects, Bonferroni multiple comparison was conducted as post-hoc analysis. For significant interactions, simple effects tests were conducted.

### *3.1. Task completion time (TIME)*

TIME was significantly affected by SIZE ($F(2,34) = 11.787$, $p < .001$) by SPACING ($F(2,34) = 12.477$, $p < .001$) and by METHOD ($F(1,17) = 34.169$, $p < .001$) (Table 1). As shown in Fig. 3, TIME appeared to decrease as SIZE increased. The SIZE of 15 mm showed the longest TIME (1.306 s), followed by 20 mm (1.162 s), and 25 mm (1.077 s). There was no significant difference in TIME for SIZEs of 20 mm and 25 mm. The SPACING of 1 mm showed the longest TIME (1.246 s), followed by 9 mm (1.153 s), and 5 mm (1.146 s). There was no significant difference in TIME for SPACINGs of 5 mm (1.146 s), and 9 mm (1.153 s). The TIME required for the METHOD of touch and push were 1.063 s and 1.3 s, respectively, which were significantly different from one another. SIZE significantly interacted with METHOD (F(2,34) = 3.547, $p < .05$). When SIZE was 15 mm, the TIME difference between touch and push was larger than the other SIZEs (See Fig. 4). An analysis of simple effects showed that the SIZE effect was significant for both the touch METHOD (F(2,34) = 4.68, $p < .05$), and the push METHOD (F(2,34) = 19.05, $p < .001$). The METHOD effect was significant for the 15-mm SIZE (F(1,17) = 62.92, $p < .001$), for the 20-mm SIZE (F(1,17) = 12.32 $p < .001$), and for the 25-mm SIZE (F(1,17) = 12.37, $p < .01$).

Table 1. ANOVA table for TIME showing only significant factors.

| Source | df | SS | MS | F | p |
|---|---|---|---|---|---|
| SIZE | 2 | 2.906 | 1.453 | 11.787 | < .001 *** |
| Subject x SIZE | 34 | 4.191 | 0.123 | - | - |
| SPACING | 2 | 0.675 | 0.337 | 12.477 | < .001 *** |
| Subject x SPACING | 34 | 0.920 | 0.027 | - | - |
| METHOD | 1 | 4.527 | 4.527 | 34.169 | < .001 *** |
| Subject x METHOD | 17 | 2.253 | 0.133 | - | - |
| SIZE x METHOD | 2 | 0.859 | 0.429 | 3.547 | 0.040 * |
| Subject x SIZE x METHOD | 34 | 4.117 | 0.121 | - | - |

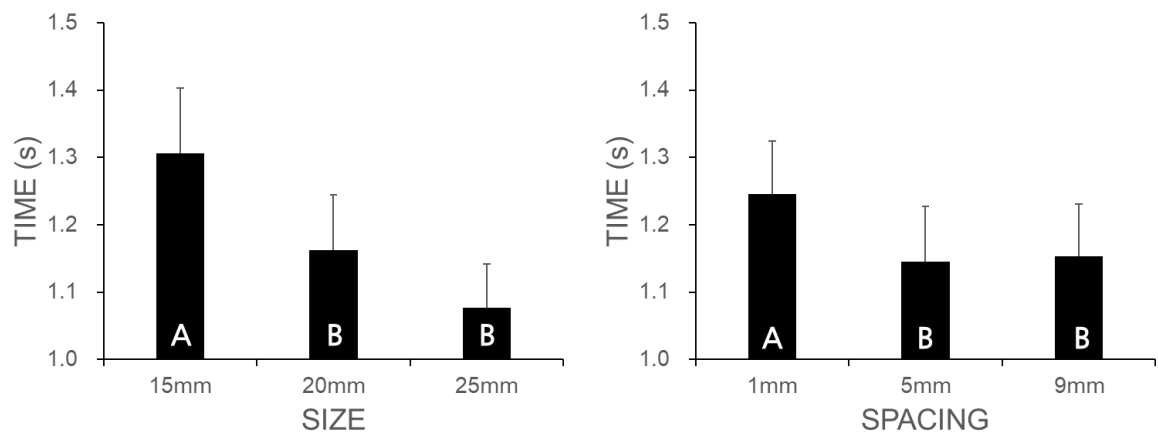

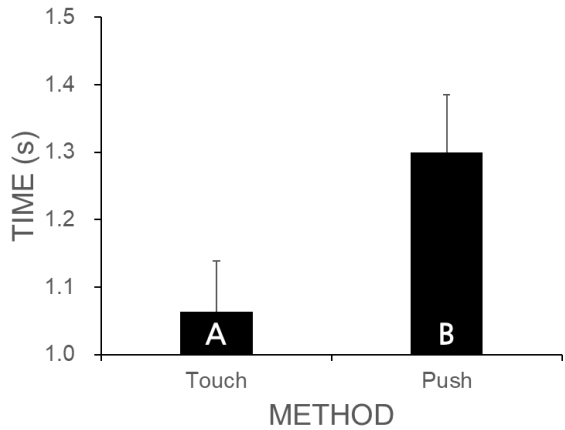

Fig. 3. Task completion time for SIZE, SPACING, and METHOD. The same alphabet characters indicate no significant difference by Bonferroni post-hoc comparison with α = .05.

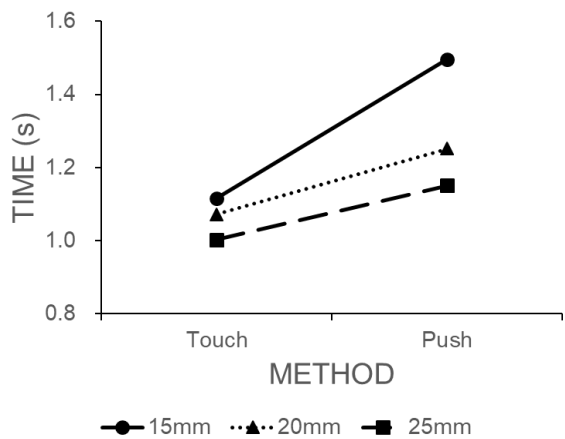

Fig. 4. Task completion time for SIZE by METHOD.

### 3.2. Number of errors (ERROR)

There was a significant effect of SIZE (F(2,34) = 19.592, $p < .001$), of SPACING (F(2,34) = 86.527, $p < .001$), and of METHOD (F(1,17) = 13.249, $p < .001$) (Table 2). As shown in Fig. 5, the number of ERRORs was the highest with 2.45 times when the SIZE was 15 mm. 20-mm SIZE (1.26 times) and 25 mm SIZE (0.92 times) were the next highest ERROR producers. There was no significant difference between SIZEs of 20 mm and 25 mm in the post-hoc analysis. In SPACING, 1-mm SPACING recorded the highest number of errors (2.92 times). The 5-mm SPACING and 9-mm SPACING recorded 1.01 times and 0.7 times, respectively. Similarly, there was no significant difference between the two SPACINGs in the post-hoc analysis. For METHOD, the number of errors was 1.96 times and 1.12 times in the touch and push method, respectively. It showed that 1.96 times is statistically different from 1.12 times.

In addition, several interaction effects were discovered (see Fig. 6). SIZE significantly interacted with SPACING ($F(4, 68) = 4.192$, $p < 0.01$). The analysis of simple effects showed that the SIZE effect was significant for 1-mm SPACING ($F(2,34) = 9.19$, $p < .001$), for the 5-mm SPACING ($F(2,34) = 14.11$ $p < .001$), and for the 9-mm SPACING ($F(2,34) = 7.48$, $p < .01$). The SPACING effect was significant for the 15-mm SIZE ($F(2,34) = 22.16$, $p < .001$), for the 20-mm SIZE ($F(2,34) = 32.44$ $p < .001$), and for the 25-mm SIZE ($F(2,34) = 9.27$, $p < .001$).

There were significant interactions between SIZE and METHOD ($F(2, 34) = 4.384$, $p < .05$). The analysis of simple effects showed that the SIZE effect was significant for both the touch METHOD ($F(2,34) = 11.49$, $p < .001$), and the push METHOD ($F(2,34) = 9.89$, $p < .001$). The METHOD effect was significant for the 15-mm SIZE ($F(1,17) = 11.62$, $p < .01$), but not for the 20-mm SIZE ($F(1,17) = 3.33$, $p = 0.071$), or the 25-mm SIZE ($F(1,17) = 1.72$, $p = 0.193$). Therefore, there was no evidence suggesting that the touch METHOD differed from the push METHOD in the number of ERRORs at 20-mm and 25-mm SIZEs.

Lastly, SPACING and METHOD interacted ($F(2,34) = 12.67$, $p < .001$). The analysis of simple effects showed that the SPACING effect was significant for both the touch METHOD ($F(2,34) = 35.64$, $p < .001$), and the push METHOD ($F(2,34) = 21.65$, $p < .001$). The METHOD effect was significant for the 1-mm SPACING ($F(1,17) = 18.39$, $p < .001$), but not for the 5-mm SPACING ($F(1,17) = 0.23$, $p = 0.63$), or the 9-mm SPACING ($F(1,17)$

= 3.01, $p$ = 0.086). This means that there was no evidence that the touch METHOD differed from the push METHOD in the number of ERRORs at 5-mm and 9-mm SPACING.

Table 2. ANOVA table for ERROR showing only significant factors.

| Source | df | SS | MS | F | p |
|---|---|---|---|---|---|
| SIZE | 2 | 140.636 | 70.318 | 19.592 | < .001 *** |
| Subject x SIZE | 34 | 122.031 | 3.589 | - | - |
| SPACING | 2 | 310.636 | 155.318 | 86.527 | < .001 *** |
| Subject x SPACING | 34 | 61.031 | 1.795 | - | - |
| METHOD | 1 | 57.086 | 57.086 | 13.249 | 0.002 ** |
| Subject x METHOD | 17 | 73.247 | 4.309 | - | - |
| SIZE x SPACING | 4 | 29.012 | 7.253 | 4.192 | 0.004 ** |
| Subject x SIZE x SPACING | 68 | 117.654 | 1.730 | - | - |
| SIZE x METHOD | 2 | 18.451 | 9.225 | 4.384 | 0.020 * |
| Subject x SIZE x METHOD | 34 | 71.549 | 2.104 | - | - |
| SPACING x METHOD | 2 | 57.080 | 28.540 | 12.670 | < .001 *** |
| Subject x SPACING x METHOD | 34 | 76.586 | 2.253 | - | - |

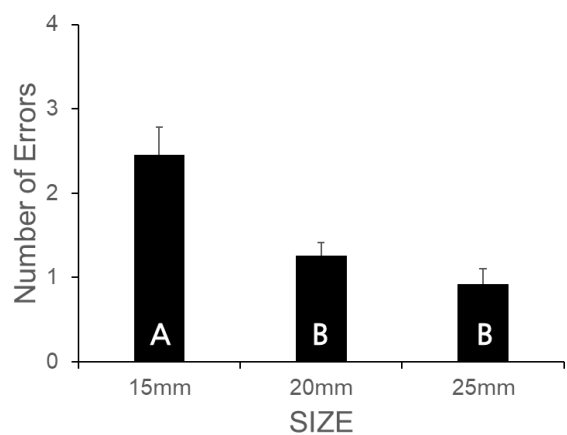
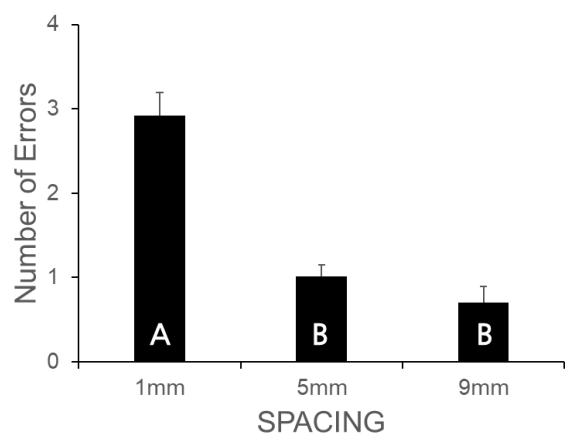

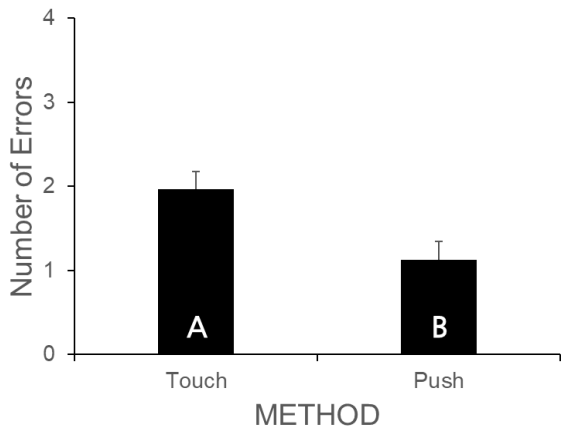

Fig. 5. Number of errors for SIZE, SPACING, and METHOD. The same alphabet characters indicate no significant difference by Bonferroni post-hoc comparison with α = .05.

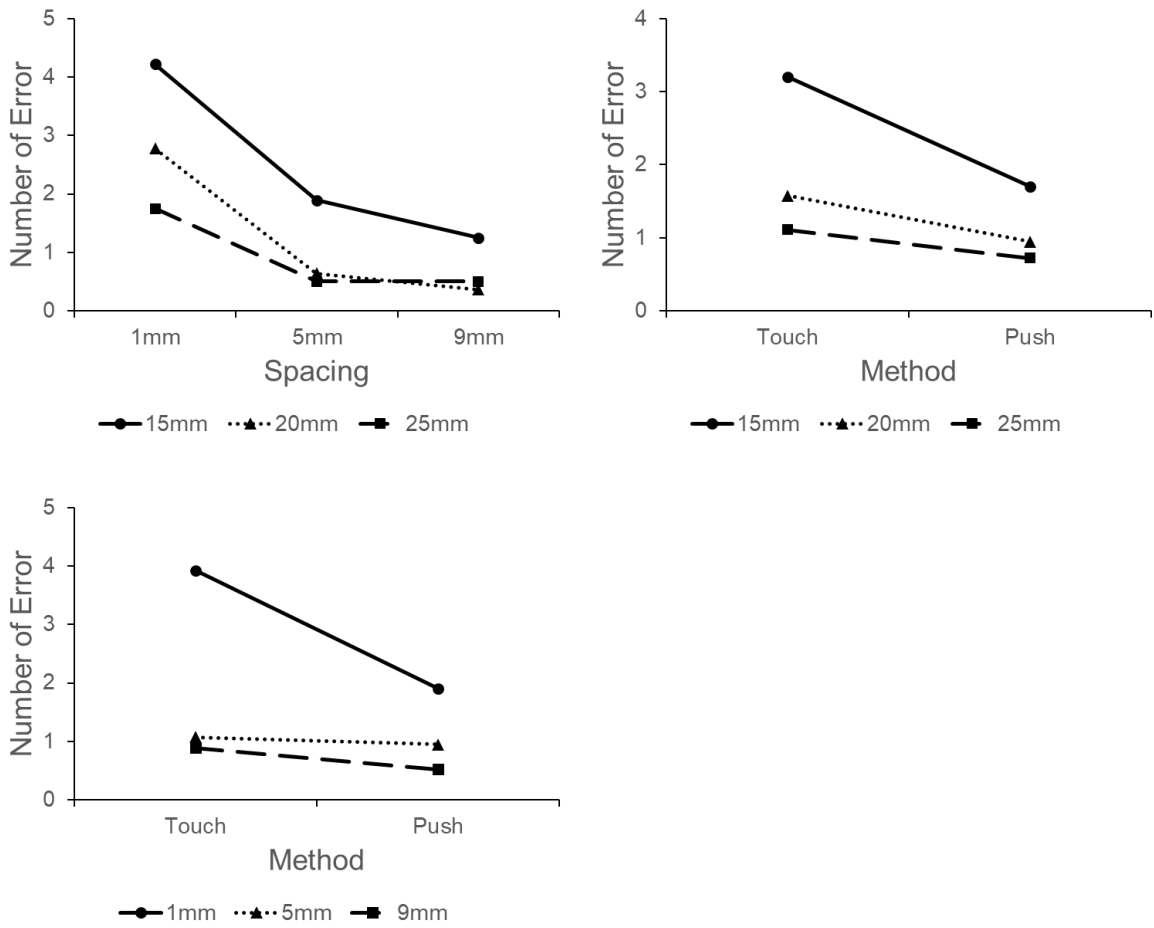

Fig. 6. Number of errors for various interactions (SIZE by SPACING, SIZE by METHOD, and SPACING by METHOD)

## 3.3. Subjective preference (PREF)

The subjective preference (i.e., PREF) was affected by SIZE (F(2,34) = 61.79, $p < .001$), SPACING (F(2,34) = 36.28, $p < .001$) (see Table 3). However, METHOD (F(1,17) = 6.51, $p = 0.13$) was not statistically significant. The PREF was the lowest at 15-mm SIZE, which was worth 54.4 points. The 20-mm SIZE and 25-mm SIZE recorded 66.9 points and 71.9 points, respectively. There existed a trend that SIZE and PREF together increased. In SPACING, 1 mm recorded 56.4 points, which was the lowest, followed by 5 mm and 9 mm with 68.2 points and 68.5 points, respectively. As a result of the post-hoc analysis, SPACINGs of 5 mm and 9 mm showed no statistically significant differences. For METHOD, the touch method received a higher score (66.1 points), than the push method (62.7 points), but the difference was not statistically significant.

Table 3. ANOVA table for PREF showing only significant factors.

| Source | df | SS | MS | F | p |
| --- | --- | --- | --- | --- | --- |
| SIZE | 2 | 17530.302 | 8765.151 | 39.269 | < .001 *** |
| Subject x SIZE | 34 | 7589.031 | 223.207 | - | - |
| SPACING | 2 | 10292.895 | 5146.448 | 22.343 | < .001 *** |
| Subject x SPACING | 34 | 7831.438 | 230.336 | - | - |

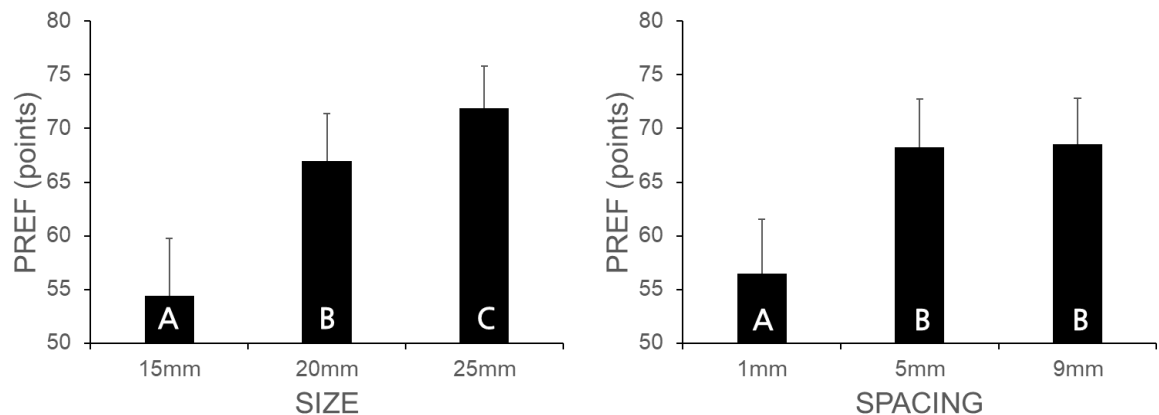

Fig. 7. Subjective preference (PREF) for SIZE and SPACING. The same alphabet characters indicate no significant difference by Bonferroni post-hoc comparison with α = .05.

## 4. Discussion

### 4.1. Button size (SIZE)

The results showed that the subjects took more time to perform the selection tasks at 15-mm SIZE compared to 20-mm SIZE and 25-mm SIZE. More ERRORs were also recorded at 15-mm SIZE. From the interview session, the subjects preferred larger SIZEs of buttons. Thus, the 15-mm SIZE for the VR HMD with a hand-motion environment was too small for utility. We developed linear regression models to understand the trend with each dependent measures (see Fig. 8). In all models, there is not enough evidence at the α = 0.05 level to conclude that there is lack of fit in the simple linear regression model. The models showed that TIME and ERROR tend to decrease as SIZE increases, and PREF tends to increase as SIZE increases. Within the experimental conditions, therefore, 25-mm SIZE is the most recommended. However, even if SIZE becomes very large, performance such as TIME and ERROR will be limited. To create a more accurate model that can explain this, it will be necessary to refine the SIZE range over 25 mm to explore the appropriate button SIZE in future research.

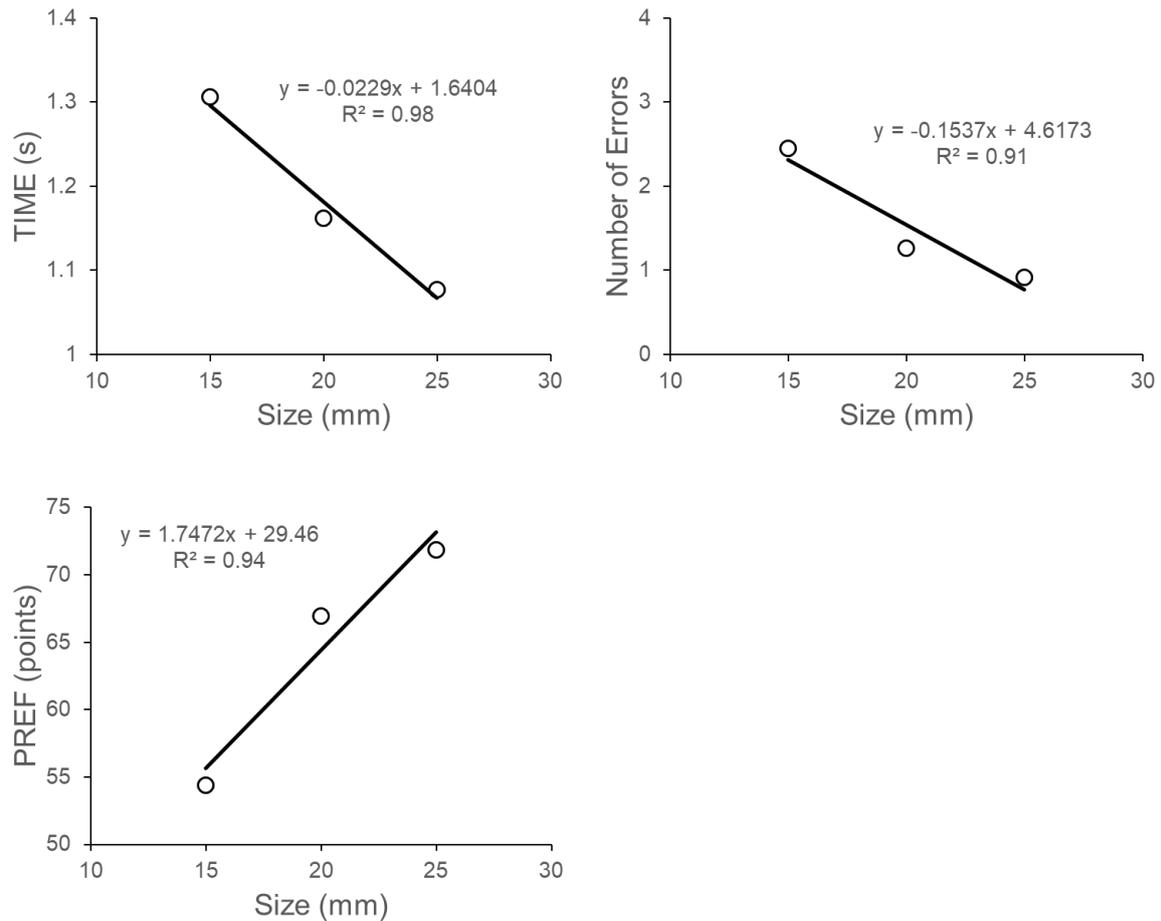

Fig. 8. Plots of each dependent measure by SIZE. Trend line was drawn using a linear model.

### 4.2. Button spacing (SPACING)

Generally, it takes longer to perform a selection task in small SPACING conditions, because subjects must be more precise in their selection process. Similar results were obtained from this experiment. In 1-mm SPACING, the longest selection task time was spent, and the highest number of errors were recorded. The subjective preference score was also the lowest. Therefore, 1-mm SPACING was not appropriate for utility. Additionally, 9-mm SPACING of the touch method with 25-mm SIZE and the push method with 15-mm SIZE and 25-mm SIZE took longer to perform the task than 5-mm SPACING. This was because too large

SPACING increases the selection range of users and leads to subjects reaching out even farther, requiring longer task completion time. Subject #3 stated, "When the SPACING increased, then I had to reach out farther, because the selection range was too spread out. My arm got tired." The subject preference showed similar results in both 5-mm SPACING and 9-mm SPACING. However, some of the cases in 9-mm SPACING were lower than those of 5-mm SPACING. We also developed regression models to understand the trend (see Fig. 9). In case of ERROR, we developed quadratic model, because there is sufficient evidence at the α = 0.05 level to conclude that there is lack of fit in the simple linear regression model. The models showed that TIME and ERROR tend to decrease as SPACING increases, and PREF tends to increases as SPACING increases. $R^2$ values of linear models for SPACING are lower than the values of models for SIZE, because the value of dependent measures at 5-mm and 9-mm SPACING were analogous unlike 20-mm and 25-mm SIZE. All of these things taken together, 1-mm SPACING is not recommended within the experimental conditions and the optimal value of SPACING is between 5 mm and 9 mm. To find out more detailed optimal SPACING, the narrower range of SPACING (e.g. 5 mm, 6 mm, 7 mm, 8 mm, and 9 mm) should be studied in the future.

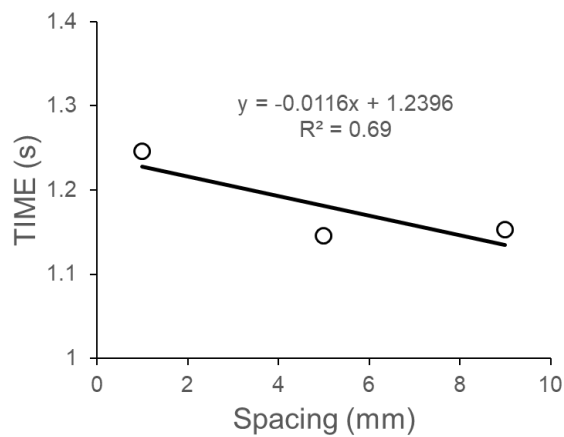
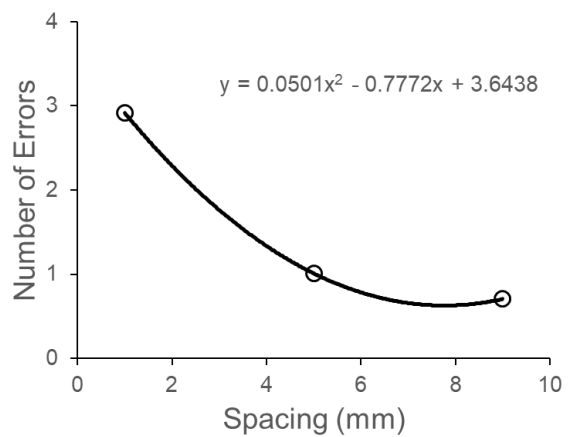

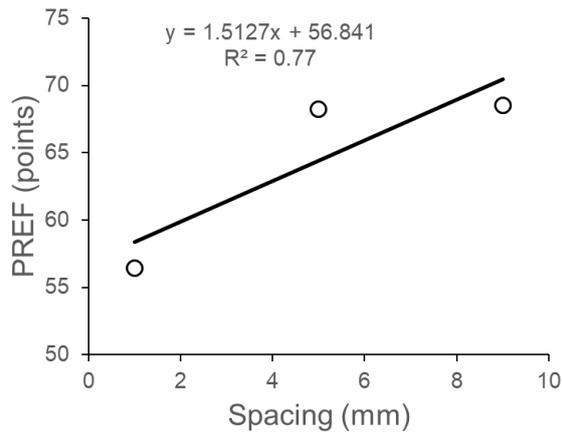

Fig. 9. Plots of each dependent measure by SPACING. Trend lines of TIME and PREF were drawn using the linear model, and trend line of ERROR was drawn using the quadratic model.

*4.3. Button depth (METHOD)*

The touch method, (depth = 0 mm), recorded a shorter task completion time compared to the push method, (depth = 15 mm). However, there were more errors in the touch method. Most subjects revealed that touching a button task finished as soon as the button was touched, which led them to quickly perform the tasks. This perhaps led to a shorter task completion time and more errors for touch buttons. However, pushing tasks required a little bit more concentration. Subject #10 said, "the buttons with the push method required more focus in performing the task. I felt I would have to put more force with the pushing method." Subject #11 shared a similar experience, "even while I was pushing the button, I felt I would have to push in the same way as the button gets pushed." In this paper, the difference of workload (i.e., touch and push method) was not investigated. However, we could infer from the collected opinions that the touch method recorded shorter task completion times because it required less workload. The lower number of errors in the push method may have occurred because of the visual feedback of the virtual buttons. Subjects could visually check and recognize the target button being pushed. If they began to push a wrong button, then they simply corrected their position to the target button before an error event was recorded. In subjective preferences, there was no significant difference between the touch and the push method. The advantages and disadvantages of the two methods obtained from the interview sessions can be summarized as follows. The touch method was faster, more convenient, and more familiar. Subjects who had used a tablet PC or an iPad were very familiar with the touch method. The disadvantage of the touch button was difficulty in recognizing when the index finger would make contact. Subject #9 said, "I was just reaching out my arm, but then a button simply got touched. I really did not mean it." However, the advantage of the push method is visual feedback. Pushing a button gave visual feedback to subjects, which allowed them to obtain presence and immersive experiences. Subject #8 said, "I feel excited when I actually see the buttons going up and down." Subject #18 also said, "when a button is being pushed, I felt I was actually pushing a button." The disadvantage of the push button was the extra concentration and effort to push, which is absent during touching. Another disadvantage was the lack of physical feedback. Interestingly, no subjects complained about the lack of physical feedback of touch-buttons. However, most subjects stated it would be better if there were physical feedback when pushing a button. There was even a case where the lack of physical feedback had a large impact in the satisfaction of the push buttons. Subject #10 said, "Virtual push buttons looked real to me, but when I found out that it gave no physical feedback, I felt very awkward and weird." In all, when designing a button for a selection task using hand interaction, a function bringing visible focus to the button being pointed to was recommended. With this function, users would be able to easily recognize which button would be touched or pushed before making contact. For systems that concentrate on an immersive experience, alarms or vibrating sensors can also be used to provide physical feedback. Electrical Muscle Stimulation is an example physical feedback, causing muscle contractions by sending small (i.e., mA scale) electrical impulses via electrodes attached to the skin (Lopes et al., 2015).

*4.4. Fitts' law analysis*

In this subsection, we introduce the result of Fitts' law analysis. Fitts' law is a well-known model in which the movement time (MT) required to perform a pointing task is linearly correlated to the index of difficulty (ID) (Fitts, 1954).

$$MT = a + b \times ID \tag{1}$$

ID can also be obtained.

$$ID = \log_2(\frac{D}{W_e} + 1) \tag{2}$$

*D* represents the distance between two targets, and $W_e$ is the effective target width, indicating the range of actual input hits around the target. It is defined as the width within 96 % of the total touch points (Mackenzie, 1992; Zhai et al., 2004; Kong and Ren, 2006). To perform the Fitts' law analysis, we extracted the time taken between task start and correct button-pressing. With standard deviation (SD) based on 2D coordination, we calculated the effective target width, ($W_e$ = 4.133 SD). We also calculated the distance, D, using an index finger trajectory extracted by experimental software. Distance values were measured in millimeter and rounded off to the nearest 10[th] digit.

The results of Fitts' analysis are shown in Fig. 10. Whereas there was some limitation in our study, in that the starting point was not controlled perfectly, the coefficient of determination ($R^2$) of our data, 0.90 for touch method and 0.93 for push method, was relatively high, compared to earlier studies on various devices. $R^2$ of Fitts' fitting was 0.83 with a mouse (Card et al., 1978), 0.86 with a trackball (Epps, 1986), and 0.55 with a touchpad (Epps, 1986). However, our results were slightly lower than 0.96 with hand (Fitts, 1954) and 0.98 with the joystick (Jagacinski and Monk, 1985).

$R^2$ was higher in the push method than in the touch. The movement time in the push method increased more than in the touch method as the ID increased, reflecting the fact that the push method had a longer task completion time.

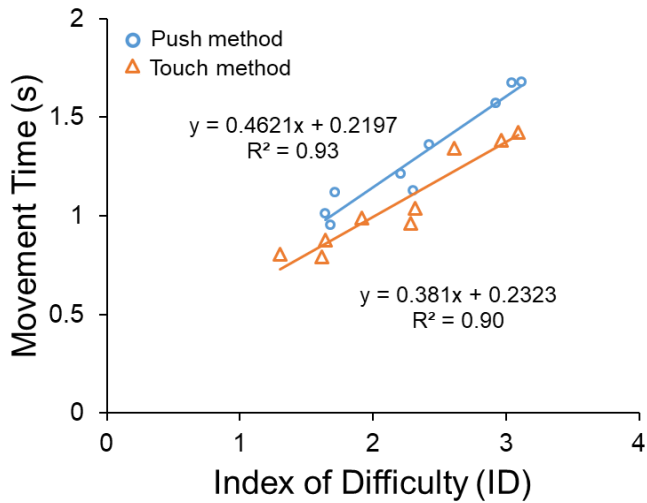

Fig. 10. Plots of Movement Time (MT) vs. Index of Difficulty (ID) for METHODs.

*4.5. Button location*

Our experimental objective was for subjects to touch or push a virtual button in 3 x 3 layout. In this section, we discuss task completion time and the number of errors, depending on the location of buttons. As Fig. 11 and Fig. 12 show, there exists an increasing trend in task completion time when the buttons were placed at the top-left corner and a decreasing trend in task completion time when the buttons were placed at the bottom-right corner. Since this experiment was held for right-handed subjects, buttons located at the bottom-right corner, which were easily approachable for right-handed subjects, showed the fastest task completion times. Alternatively, a longer duration of time was required for buttons located at the top-left corner. From the error perspective, the center button (i.e., Button #5) recorded the highest number of errors: 4.67 times. The reason can be inferred from the fact that this button was surrounded by all the other buttons. Subjects had difficulties touching or pushing the center button without inadvertently touching or pushing the surrounding buttons. High number of errors were recorded as 4.33 and 4.61 times at the top-left corner buttons (i.e., Buttons #1 and #4, respectively). This indicates that touching or pushing buttons located at the top-left corner, where the buttons were the farthest, was very difficult for right-handed subjects.

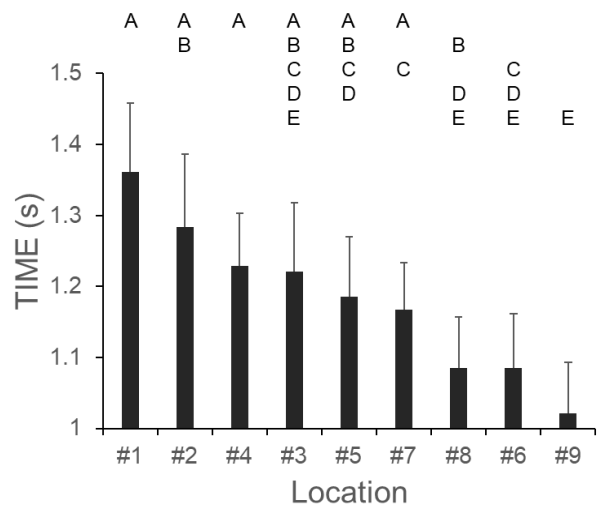

| | | |
|---|---|---|
| **#1** 1.36 | **#2** 1.28 | **#3** 1.22 |
| **#4** 1.23 | **#5** 1.19 | **#6** 1.09 |
| **#7** 1.17 | **#8** 1.09 | **#9** 1.02 |

Fig. 11. Task completion time for button locations. The same alphabet characters indicate no significant difference by Bonferroni post-hoc comparison with $\alpha = .05$.

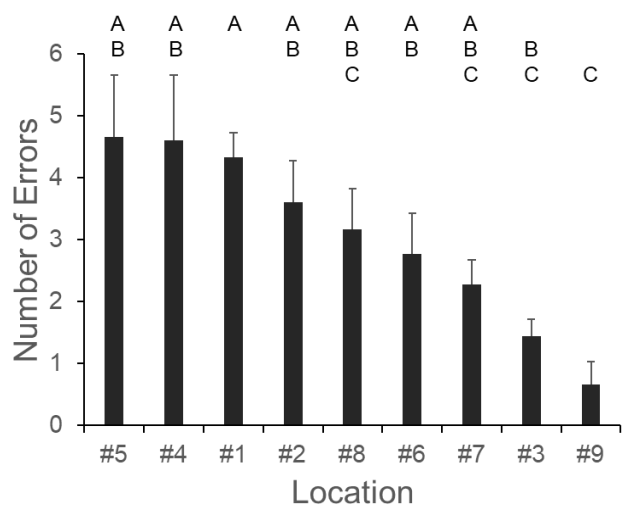

| | | |
|---|---|---|
| **#1** 4.33 | **#2** 3.61 | **#3** 1.44 |
| **#4** 4.61 | **#5** 4.67 | **#6** 2.78 |
| **#7** 2.28 | **#8** 3.17 | **#9** 0.67 |

Fig. 12. The number of errors for location of buttons. The same alphabet characters indicate no significant difference by Bonferroni post-hoc comparison with α = .05.

## 4.6. Gender

In this subsection, we explore the gender effect on this experiment. In TIME, females took longer time (1.298 s) to complete the tasks than male subjects (1.065 s). However, the difference was not statistically significant ($F(1,17) = 2.444$, $p = 0.138$). There existed no other factors to noticeably affect the TIME factor. In ERROR, male subjects recorded higher numbers of errors, (1.932 times) compared to female subjects (1.154 times), where the result was significant ($F(1,17) = 5.38$, $p < .05$). There existed an interaction effect (see Fig. 13) between the SIZE and the gender effect ($F(2,34) = 6.19$, $p < .01$). From the simple effect analysis, the gender effect was significant at a SIZE of 15 mm ($F(1,17) = 19.10$, $p < .001$) and 20 mm ($F(1,17) = 5.82$, $p < .05$). However, the gender effect at 25-mm SIZE showed no significant effect. The effect of SIZE showed that there existed significant differences between the male group ($F(2,34) = 26.23$, $p < .001$), and the female group ($F(2,34) = 4.21$, $p < .05$). For PREF, male subjects tended to give relatively higher subjective scores (69.3 points) than females, (59.5 points), where the difference was not significant ($F(1,17) = 1.221$, $p = 0.286$). The gender effect had an interaction effect with METHOD ($F(1,17) = 7.69$, $p < .05$) (see Fig. 14). From the simple effect analysis, the gender effect showed a significant difference in touch METHOD ($F(1,17) = 9.31$, $p < .01$), and push METHOD ($F(1,17) = 10.35$, $p < .01$). The METHOD effect was significantly important to female groups ($F(1,17) = 5.62$, $p < .05$), whereas it was not so for the male group ($F(1,17) = 0.22$, $p = 0.64$). Thus, it can be inferred that females tend to prefer the touch method over the push method. However, males do not care whether the method was push or touch. This clearly describes that there exist the differences based on gender. In the debriefing session, there were different types of replies between the male and female groups. Five out of nine females negatively replied that the push method was difficult, because they had to exert more force. Two positively replied that the push button was visually amusing, because they could see that the button was being pushed. However, most males, six of nine, positively replied that the presence was high, owing to the visual effect, and felt happy about it. Only two replied that they experienced difficulties, because they had to focus more on the push METHOD. Additionally, they provided opinions that it was difficult to concentrate because there was no tactile feedback during touch.

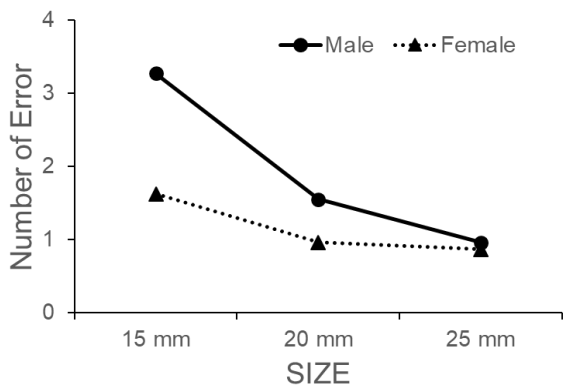

Fig. 13. Number of errors for interaction (SIZE by GENDER)

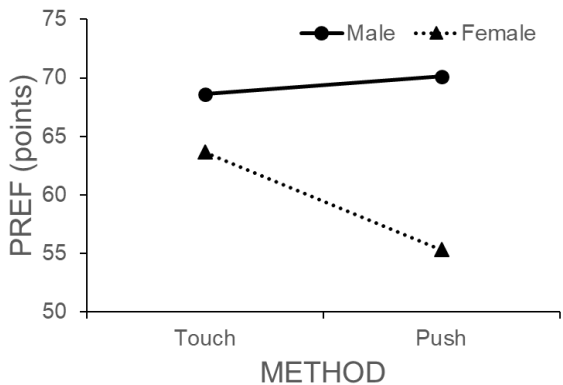

Fig. 14. Subjective preference score (PREF) for interaction (METHOD by GENDER)

## 5. Conclusion

In this study, the effect of button size, spacing, and depth on the usability of VR HMDs with hand interaction was investigated. Within the experimental conditions, 25 mm was the best, in terms of SIZE. The spacing was optimal at between 5 mm and 9 mm, and the excessively large spacing increased the fatigue and caused aesthetic defects, affecting the subjective preference negatively. There was no significant difference in subjective preference between the two buttons, per the differences of depth. However, the touch method was faster than the push method. Yet, the number of errors was higher. It is important for the VR HMD to create immersive experiences with various feedback methods, given that the subjects required visual feedback for the touch method and requested tactile feedback for the push method. The results of this experiment are expected to be useful when designing VR buttons to control via hand interaction. Further research should explore various new experimental conditions. In this study, owing to the lack of research on button design in the VR environments, we substituted the experimental condition of other similar areas. This study is a starting point, and various new studies should be performed. Additionally, considering the scalability and applicability of VR HMDs, further investigation and experimentation into subject age is strongly recommended.

**Acknowledgement**

This research was supported by the National Research Foundation of Korea (NRF) grant funded by the Korea government (MSIP) (No. NRF-2016R1A2B2011158)

# References

Bababekova, Y., Rosenfield, M., Hue, J. E., & Huang, R. R. (2011). Font size and viewing distance of handheld smart phones. *Optometry & Vision Science*, 88(7), 795-797.

Bertram, J., Moskaliuk, J., & Cress, U. (2015). Virtual training: Making reality work?. *Computers in Human Behavior*, *43*, 284-292.

Burdea, G. C., & Coiffet, P. (2003). *Virtual reality technology* (Vol. 1). Wiley.

Card, S. K., English, W. K., & Burr, B. J. (1978). Evaluation of mouse, rate-controlled isometric joystick, step keys, and text keys for text selection on a CRT. *Ergonomics*, *21*(8), 601-613.

Colle, H. A., & Hiszem, K. J. (2004). Standing at a kiosk: Effects of key size and spacing on touch screen numeric keypad performance and user preference. *Ergonomics*, *47*(13), 1406-1423.

Epps, B. W. (1986, September). Comparison of six cursor control devices based on Fitts' law models. In *Proceedings of the Human Factors Society Annual Meeting* (Vol. 30, No. 4, pp. 327-331). Sage CA: Los Angeles, CA: SAGE Publications.

Fitts, P. M. (1954). The information capacity of the human motor system in controlling the amplitude of movement. *Journal of experimental psychology*, *47*(6), 381.

Hofmann, M., Brger, R., Frost, N., Karremann, J., Keller-Bacher, J., Kraft, S., ... & Steinicke, F. (2013). Comparing 3d interaction performance in comfortable and uncomfortable regions. In *Proceedings of the GI-Workshop VR/AR* (pp. 3-14).

IJsselsteijn, W., Van Den Hoogen, W., Klimmt, C., De Kort, Y., Lindley, C., Mathiak, K., ... & Vorderer, P. (2008, August). Measuring the experience of digital game enjoyment. In *Proceedings of Measuring Behavior* (pp. 88-89). Wageningen, Netherlands: Noldus Information Tecnology.

Jagacinski, R. J., & Monk, D. L. (1985). Fitts' Law in Two dimensions with hand and head movements. *Journal of Motor Behavior*, *17*(1), 77-95.

Kim, H., Kwon, S., Heo, J., Lee, H., & Chung, M. K. (2014). The effect of touch-key size on the usability of In-Vehicle Information Systems and driving safety during simulated driving. *Applied ergonomics*, 45(3), 379-388.

Kim, S., & Kim, G. J. (2004). Using keyboards with head mounted displays. In *Proceedings of the 2004 ACM SIGGRAPH international conference on Virtual Reality continuum and its applications in industry* (pp. 336-343). ACM.

Kong, J., & Ren, X. (2006). Calculation of effective target width and its effects on pointing tasks. *Information and Media Technologies*, *1*(2), 1057-1059.

Lopes, P., Ion, A., & Baudisch, P. (2015). Impacto: Simulating physical impact by combining tactile stimulation with electrical muscle stimulation. In *Proceedings of the 28th Annual ACM Symposium on User Interface Software & Technology* (pp. 11-19). ACM.

Lawson, G., Salanitri, D., & Waterfield, B. (2016). Future directions for the development of virtual reality within an automotive manufacturer. *Applied ergonomics*, 53, 323-330.

Lubos, Paul, Gerd Bruder, and Frank Steinicke. "Analysis of direct selection in head-mounted display environments." 3D User Interfaces (3DUI), 2014 IEEE Symposium on. IEEE, 2014.

MacKenzie, I. S. (1992). Fitts' law as a research and design tool in human-computer interaction. *Human-computer interaction*, *7*(1), 91-139.

Martel, E., & Muldner, K. (2017). Controlling VR games: control schemes and the player experience. *Entertainment Computing*, *21*, 19-31.

Mazuryk, T., & Gervautz, M. (1996). Virtual reality-history, applications, technology and future.

Mine, M. R., Brooks Jr, F. P., & Sequin, C. H. (1997, August). Moving objects in space: exploiting proprioception in virtual-environment interaction. In *Proceedings of the 24th annual conference on Computer graphics and interactive techniques* (pp. 19-26). ACM Press/Addison-Wesley Publishing Co.

Parhi, P., Karlson, A. K., & Bederson, B. B. (2006, September). Target size study for one-handed thumb use on small touchscreen devices. In *Proceedings of the 8th conference on Human-computer interaction with mobile devices and services* (pp. 203-210). ACM.

Park, Y. S., & Han, S. H. (2010). Touch key design for one-handed thumb interaction with a mobile phone: Effects of touch key size and touch key location. *International journal of industrial ergonomics*, *40*(1), 68-76.

Park, Y. S., (2010). User Interface Design for One-handed Touch Interaction with Mobile Devices, Master's thesis. Pohang University of Science and Technology, 2010.

Rose, T., Nam, C. S., & Chen, K. B. (2018). Immersion of virtual reality for rehabilitation-Review. *Applied ergonomics*, 69, 153-161.

Scott, B., & Conzola, V. (1997, October). Designing touch screen numeric keypads: Effects of finger size, key size, and key spacing. In *Proceedings of the Human Factors and Ergonomics Society Annual Meeting* (Vol. 41, No. 1, pp. 360-364). Sage CA: Los Angeles, CA: SAGE Publications.

Sekuler, R., & Blake, R. (1985). Perception. Alfred A. *Kopf NY*.

Sun, H. M., Li, S. P., Zhu, Y. Q., & Hsiao, B. (2015). The effect of user's perceived presence and promotion focus on usability for interacting in virtual environments. *Applied ergonomics*, 50, 126-132.

Sutherland, I.E. The ultimate display. In: Proceedings of IFIP Congress. 1965: 506-508.

Zhai, S., Kong, J., & Ren, X. (2004). Speed–accuracy tradeoff in Fitts' law tasks—on the equivalency of actual and nominal pointing precision. *International journal of human-computer studies*, *61*(6), 823-856.

Zhang, H. (2017). Head-mounted display-based intuitive virtual reality training system for the mining industry. *International Journal of Mining Science and Technology*.